%% file: main.tex
\documentclass[ALICE,manyauthors]{cernphprep}
\usepackage[comma,square,numbers,sort&compress]{natbib}
\usepackage{hyperref}
\usepackage{lineno}
\usepackage{xspace}
\usepackage{booktabs}
\usepackage{multirow}
\usepackage{hyperref}
\usepackage{cleveref}
\usepackage{ulem}

\usepackage[T1]{fontenc}
\usepackage{orcidlink} 

\begin{document}
\input{commands.tex}

\begin{titlepage}
\PHyear{2026}       
\PHnumber{030}      
\PHdate{11 Feb}  

\title{Deuteron coalescence probability in jets in \\~\pPb collisions at \fivenn}
\ShortTitle{Deuteron coalescence probability in jets in \pPb}  

\Collaboration{ALICE Collaboration\thanks{See Appendix~\ref{app:collab} for the list of collaboration members}}
\ShortAuthor{ALICE Collaboration} 

\begin{abstract}
This article presents the first measurement of the (anti)deuteron and antiproton transverse-momentum distributions, the deuteron coalescence parameter, and the ratio between the yields of deuteron and proton in and out of jets in \pPb collisions at \fivenn. Three regions are used to study the jet-correlated production: Toward, Away, and Transverse to the jet axis, which direction is approximated by the leading particle in the event with $ p_{\mathrm{T}}^{\mathrm{lead}} >$ 5~\GeVc. The jet contribution is obtained from the Toward region by subtracting the underlying event, which dominates the Transverse region. In \pPb collisions, an enhancement of the coalescence parameter $B_{\mathrm{2}}$ is observed in jets compared to the underlying event. The enhancement is larger than the one observed in a similar analysis already in pp collisions. The results are compared with predictions from PYTHIA 8.314 using the Angantyr model with a deuteron production model based on ordinary nuclear reactions. The model is able to qualitatively reproduce the large enhancement of the coalescence parameter in jets with respect to the out-of-jets one. The results are further compared in detail with similar ones from previous studies for \pp collisions at \thirtheen.

\end{abstract}
\end{titlepage}

\setcounter{page}{2} 


\section{Introduction} 
\label{sect:intro}
The production of light (anti)nuclei has been extensively studied at different collision energies, ranging from AGS~\cite{AGS_1, AGS_2, AGS_3, AGS_4}, to the SPS~\cite{SPS}, RHIC~\cite{RHIC_1, RHIC_2, RHIC_3, RHIC_4, RHIC_5, RHIC_6} and the LHC~\cite{ALICE_nuclei_antinuclei, ALICE_nuclei_pp_PbPb, ALICE_deuteron_multiplicity_pp_7_TeV, ALICE_deuteron_elliptic_flow_PbPb_2_TeV, ALICE_nuclei_pp_run1, ALICE_helium4_PbPb_2TeV, ALICE_nuclei_multiplicity_pPb_5TeV, ALICE_triton_helium3_pPb_5TeV, ALICE_deuteron_pp_13TeV, ALICE_pp_HM, ALICE_deuteron_jets, ALICE_nuclei_pp_5TeV, ALICE_hypertrition_pPb_5TeV, ALICE_elliptic_triangular_flow_deuteron_PbPb_5TeV, ALICE_elliptic_flow_helium, ALICE_pp_B2_jet, ALICE_nuclei_pPb_8TeV, ALICE_deuteron_fluctuation, ALICE_nuclei_PbPb_5TeV}. Despite the abundance of experimental results, the production mechanism of light (anti)nuclei is still unclear. The experimental data are typically described by two different models: the statistical hadronization model (SHM) and the baryon coalescence model. In the SHM~\cite{SHM_1, SHM_2, SHM_3, SHM_4, SHM_5, SHM_6, SHM_7, SHM_8}, hadrons are emitted by a source in local thermal and hadrochemical equilibrium, with the species abundance fixed at chemical freeze-out. In nucleus-nucleus collisions, given that the quantum numbers are conserved on average in all the system, the predicted hadron yields are described using the grand canonical ensemble~\cite{SHM_5}, while in proton-proton and proton-nucleus collisions, since the exact conservation of the quantum numbers is required only in the correlation volume ($V_C$), the canonical approach is used~\cite{SHM_8}.  In the coalescence model~\cite{coal_1, coal_2, coal_3, coal_4, coal_5, coal_6, coal_7, coal_TUM}  nuclei are formed from the coalescence of protons and neutrons that are close in phase space. In a simpler approach~\cite{coal_2} only the momentum correlations are considered, while in more advanced implementations~\cite{coal_3, coal_TUM} the coalescence probability is calculated considering the overlap of the wave function of the individual baryons and the Wigner function of the bound state. Despite the use of the two different models, it is important to note that they do not necessarily exclude each other. In fact, the SHM only predicts the overall particle yields, not the momentum distributions. Moreover, the SHM is a macroscopic model, hence a microscopic description on the hadron formation mechanism is not discussed. The baryon coalescence model, instead, provides such microscopic description in its formulation.

Another interesting field of application concerning the study of light (anti)nuclei produced at colliders is related to indirect dark-matter searches. In fact, different space-based experiments, such as AMS-02~\cite{AMS} and GAPS~\cite{GAPS}, are actively searching the signal originating from light antinuclei in cosmic rays~\cite{space_1, space_2, space_3}. In particular, the detected flux of such particles must be correctly evaluated, in order to understand their source. In the current model, antinuclei can be produced in interactions between high-energy cosmic rays and the interstellar medium, or from annihilation of dark matter. Since the former constitutes a large background for dark-matter searches, precise estimate of the antinucleus production are critical, and collider experiments can provide valuable constraints. In fact, the measurements obtained in collider experiments not only give insights in the antinucleus production mechanism, but they can also provide information on the antinucleus cross sections. Novel results from ALICE evaluate the inelastic cross section of antideuteron~\cite{ALICE_deuteron_cross_section}, antitriton~\cite{ALICE_triton_cross_section}, and anti$^3\mathrm{He}$~\cite{ALICE_helium_cross_section} interactions with nuclei, also discussing the implications for the astrophysical searches of dark matter in the case of anti$^3\mathrm{He}$. 

The current knowledge on the light (anti)nuclei production mechanism is still incomplete, hence innovative observables to test the models' predictions are needed. In particular, to further constrain the coalescence model, the production mechanism of light (anti)nuclei can be studied in and out of jets, as already done in Refs.~\cite{ALICE_pp_B2_jet, coal_IRC}. Jets are defined as collimated emission of hadrons produced by the hadronization of energetic quarks and gluons generated in high-$Q^2$ interactions.  Due to the collimation, in general, hadrons in jets are closer in phase space than hadrons out of jets~\cite{ALICE_radial_jet_profile}. Hence, according to the coalescence model, an enhancement of the coalescence probability in jets is expected with respect to the same quantity out of jets, where the particles are not as strongly correlated.  For this reason it is interesting to study the difference of the light (anti)nuclei production in and out of jets in small systems, such as \pp and \pPb, since the nucleons are already closer in phase space, but the two systems differ in other aspects, such as spatial dimension and multiplicity of charged particles. Specifically, the source size, i.e. the region that emits hadrons with similar momenta, that can be measured with two-particle correlation measurements through the femtoscopy method, in \pPb is slightly larger with respect to the \pp case. In this case the coalescence model predicts a smaller coalescence probability out of jets in \pPb collisions with respect to the same quantity evaluated in \pp collisions~\cite{coal_3, coal_pPb_pred}. For the in-jet case, instead, an enhancement of the coalescence probability with respect to the out of jet case is predicted, due to the fact that hadrons are closer in phase space. Another aspect that can be investigated is the jet hadrochemistry. Based on current knowledge, it should be independent of the collision system from which the jets originate~\cite{jet_review_sapeta}. Hence, similar results on particle-species compositions in jets in different collision systems are expected. If, instead, the jet hadrochemistry depends on the collision systems, differences in particle-species composition and distributions in jets could arise, with consequences on the production of light (anti)nuclei and other particles~\cite{ALICE_strangeness_jet_1, ALICE_strangeness_jet_2}. In particular, for the light (anti)nuclei, the nuclei-to-proton yield ratio could give insights in the hadrochemistry.

In this article, (anti)deuteron and antiproton production is studied in and out of jets. Unlike for antiprotons, for deuterons both matter and antimatter particles are used. This choice maximizes the collected signal for the (anti)deuterons, since the production of such particles is suppressed of a factor of about $10^3$ with respect to (anti)protons, while it ensures that the antiproton data sample almost exclusively consists of primary particles (except for a small fraction of secondary particles due to weak decays), since the contribution of secondary particles created in the interaction with the detector is negligible for the antimatter case. It is important to note that the results of this analysis may depend on how the jets are defined, i.e. whether a trigger particle of a given \pt or the full \pt of the jet is considered. For this reason, the same jet definition used in pp collisions is applied to allow for a comparison with the published result~\cite{ALICE_pp_B2_jet}. In each event the leading particle, that is used as a proxy for the jet axis, is defined as the particle with the highest transverse momentum, which also satisfies the condition $p_{\mathrm{T}}^{\mathrm{lead}} >$ 5~\GeVc, where $p_{\mathrm{T}}^{\mathrm{lead}}$ is the transverse momentum of the leading particle, as reported in Ref.~\cite{ALICE_pPb_RT}. The transverse-momentum distributions are measured in three different azimuthal regions defined relative to the leading particle by $\Delta\varphi = \lvert \varphi - \varphi_{\mathrm{lead}}\rvert$, where $\varphi$ is the azimuthal angle of the particle and $\varphi_{\mathrm{lead}}$ refers to the direction of the leading track. The obtained regions, all 120\degree wide, are the same as the ones defined in similar analyses~\cite{ALICE_pp_RT, ALICE_pPb_RT}: the region around the jet axis (Toward, $\lvert\Delta\varphi\rvert < 60\degree$), the region back-to-back to the jet axis (Away, $\lvert\Delta\varphi\rvert > 120\degree$) and the region transverse to both of them (Transverse, $60\degree < \lvert\Delta\varphi\rvert < 120\degree$). The Transverse region is dominated by the underlying event, while the Toward and Away regions, in addition to the underlying event, have a contribution from the leading and recoil jet, respectively. The in-jet transverse-momentum distributions are hence obtained by  subtracting the distributions in the Transverse region from the corresponding ones in the Toward region. 

The coalescence parameter, which quantifies the coalescence probability~\cite{coal_2}, is defined as:
\begin{equation}
\centering
B_2 = \Biggl(\dfrac{1}{(2\pi/3)p_{\mathrm{T}}^{\mathrm{d}}}\dfrac{1}{N_{\mathrm{evt, trig}}}\dfrac{\mathrm{d}^{2}N_{\mathrm{d}}}{\mathrm{d}y\mathrm{d}p_{\mathrm{T}}^{\mathrm{d}}}\Biggr)\Bigg/\Biggl(\dfrac{1}{(2\pi/3)p_{\mathrm{T}}^{\mathrm{p}}}\dfrac{1}{N_{\mathrm{evt, trig}}}\dfrac{\mathrm{d}^{2}N_{\mathrm{p}}}{\mathrm{d}y\mathrm{d}p_{\mathrm{T}}^{\mathrm{p}}}\Biggr)^2 , 
\label{eqn:coalescence_parameter}
\end{equation}
where the labels d and p indicate the two different species, $p_{\mathrm{T}}^{\mathrm{p}} = p_{\mathrm{T}}^{\mathrm{d}}/2$, with the assumption that the neutron and proton momentum distributions are the same since they belong to the same isospin doublet, and $N_{\mathrm{evt, trig}}$ is the number of events with a leading particle that satisfies the trigger condition. The number of events and triggers in this analysis are the same. Unlike in previous analyses~\cite{ALICE_nuclei_pp_PbPb, ALICE_deuteron_multiplicity_pp_7_TeV, ALICE_deuteron_elliptic_flow_PbPb_2_TeV, ALICE_nuclei_pp_run1, ALICE_nuclei_multiplicity_pPb_5TeV, ALICE_triton_helium3_pPb_5TeV, ALICE_deuteron_pp_13TeV, ALICE_pp_HM, ALICE_nuclei_pp_5TeV, ALICE_nuclei_pPb_8TeV, ALICE_nuclei_PbPb_5TeV}, the (anti)deuteron and antiproton production in this analysis is studied in three regions of phase space with respect to the jet axis, which leads to an additional factor of 1/3. This factor is the same in both the in and out of jets cases, since the two regions have the same width in $\varphi$ 2$\pi$/3.
 
\section{ALICE detector and data samples}
\label{sect:ALICE}
The ALICE apparatus is ideally suited to study light (anti)nuclei production because of its excellent tracking performance and particle-identification capabilities in a broad range of transverse momentum. A detailed description of the ALICE subsystems and their performance can be found in Refs.~\cite{ALICE_description_1, ALICE_description_2} and references therein. 

The detectors used in this analysis, which are briefly described below in their state during the Run 2 data taking, are the Inner Tracking System (\ITS), the Time Projection Chamber (\TPC), and the Time-Of-Flight (\TOF) detector. These detectors are located inside a solenoidal magnet with a field strength of \textit{B} = 0.5~T. 

The \ITS~\cite{ALICE_description_1} is used to reconstruct the primary vertex of the collision and the secondary vertices from weak decays. It covers the full azimuthal angle and the pseudorapidity interval \etarange 0.9 and is composed of six cylindrical layers of silicon detectors, with a radial distance ranging from 3.9~cm to 43~cm. Three different technologies are used: the two innermost layers are Silicon Pixel Detectors (SPD), followed by two layers of Silicon Drift Detectors (SDD), and the two outermost layers are Silicon Strip Detectors (SSD). 

The \TPC~\cite{ALICE_TPC} is the main ALICE tracking detector. It is a cylindrical drift chamber with a length of 500 cm in the beam direction and a radius ranging from 85 to 250 cm, covering the same pseudorapidity interval of the ITS. The vessel is filled with a gas mixture of 90\% Ar and 10\% $\mathrm{CO}_{\mathrm{2}}$ at atmospheric pressure. The cage is closed with two endcaps made of Multi-Wire Proportional Chambers (MWPCs). The particle identification is obtained by measuring the specific energy loss (\dEdx) in the gas, while the trajectory is reconstructed using up to 159 space points reconstructed in the TPC volume.

The \TOF detector~\cite{ALICE_TOF} is formed by Multi-gap Resistive Plate Chambers (MRPCs) located at a radial distance of 380~cm from the beam axis. It measures the arrival time of charged particles relative to the event collision time provided by the \TOF itself and by the T0 detector~\cite{ALICE_T0}. The T0 detector consists of two arrays of Cherenkov counters (T0A and T0C) placed at opposite sides of the interaction point, covering the pseudorapidity regions 4.6 $< \eta <$ 4.9 and $- 3.3 < \eta < -3.0$. 

The \VZERO detector~\cite{ALICE_V0} is used to define the minimum bias trigger, and to separate beam-beam interactions from background events such as beam-gas interactions. It is formed by two plastic scintillator arrays (\VZEROA and \VZEROC) placed at both sides of the interaction point and covering the pseudorapidity regions $2.8 < \eta < 5.1$ and $- 3.7 < \eta < - 1.7$. 

A data sample collected by ALICE in 2016 during the LHC \pPb campaign at \fivenn is used in this analysis. The events are selected with a minimum bias trigger, where the signals in both \VZERO scintillators must be synchronous with the beam crossing time defined by the LHC clock. Events with multiple vertices identified with the SPD are tagged as pile-up and removed from the analysis. To ensure a uniform detector response, only events with a reconstructed vertex position along the beam axis within 10 cm from the nominal interaction point are considered.  With these requirements, a data sample of about 310 million minimum bias events is selected. Since the energy per nucleon of the proton beam is higher with respect to the Pb beam, the midrapidity axis in the center-of-mass frame is shifted with respect to the laboratory frame by -0.465 units of rapidity in the direction of the proton beam~\cite{ALICE_triton_helium3_pPb_5TeV}, that travels toward the negative z-direction. For this reason, the rapidity of each track in the laboratory system needs to be transformed into the centre-of-mass system applying the following formula: 
\begin{equation}
    \centering
    \ycms = \ylab + 0.465.
    \label{eqn:rapidity_shift}
\end{equation} 
After applying such a shift to all the particles, only the ones with $-0.07 < \ycms < 0.93$ are selected. To identify events that contain a jet, only events with at least one charged primary particle with $ p_{\mathrm{T}}^{\mathrm{lead}} >$ 5~\GeVc are considered. The final data sample corresponds to approximately 2.4\% of the minimum-bias \pPb collisions.

\section{Analysis procedure}
\label{sec:analysis}
In this section the analysis procedure is reported, including the track selection criteria, particle identification and signal extraction, corrections derived from Monte Carlo (MC) simulations, and the evaluation of the systematic uncertainties. 

\subsection{Track selection and particle identification}
\label{subsec:pid}
Antiproton and (anti)deuteron candidates are selected among charged-particle tracks reconstructed in the \ITS and \TPC in the kinematic range $\lvert\eta\rvert<0.8$. Reconstructed tracks are required to fulfill a set of quality criteria. The minimum number of clusters required in the ITS must be higher than two or four for the (anti)deuteron and antiproton, respectively, with at least one cluster located in one of the two innermost ITS layers in both cases. The minimum number of clusters in the TPC must be 50 for the (anti)deuteron and 60 for antiproton. To ensure accurate reconstruction of the fitted tracks, the $\chi^2$ per degree of freedom is required to be less than four for the \TPC for both species. The contribution of secondary nuclei is reduced by requiring that the distance of closest approach (DCA) to the primary vertex in the transverse plane ($\mathrm{DCA}_{xy}$) and longitudinal direction ($\mathrm{DCA}_{z}$) is lower than, respectively, 0.1 and 1.0~cm for both species. The leading particle is defined as the track with the highest transverse momentum among those fulfilling the same quality criteria used in previous analyses ~\cite{ALICE_pp_RT, ALICE_pp_B2_jet, ALICE_pPb_RT}. 

In the three azimuthal regions, the same procedure for signal extraction is applied. Antiprotons with $0.3 < \pt <  0.7$~\GeVc and (anti)deuterons with $0.6 < \pt < 1.2$~\GeVc are identified by requiring that the specific energy loss \dEdx measured by the \TPC~\cite{ALICE_TPC} is within 3$\sigma$ of the expected value predicted by the Bethe-Bloch parametrization. For (anti)deuterons between $1.0 < \pt < 1.2$~\GeVc,  the contamination from (anti)protons is fitted with a Gaussian function and subtracted. At higher transverse momenta, i.e. $0.7 < \pt <  1.5$~\GeVc for antiprotons and $1.2 < \pt <  3.0$~\GeVc for (anti)deuterons, the particle identification is performed using the \TPC and \TOF detectors, where the \TOF time-of-flight distributions~\cite{ALICE_TOF} are analyzed for the particles that satisfied the \TPC pre-selection criteria,  $\lvert n\sigma_{\mathrm{TPC}}\rvert <  3$, defined as $n\sigma_{\mathrm{TPC}} = \left(\dEdx - \langle \dEdx\rangle_{i}\right)/\sigma_{\mathrm{TPC, i}}$, where $\dEdx$ is the measured signal in the TPC, $\langle\dEdx\rangle_i$ is the expected signal for the particle species $i$ and $\sigma_{\mathrm{TPC,i}}$ is the detector resolution for that species.  The signal is extracted from a fit of the $n\sigma_{\mathrm{TOF}} = (\Delta t-\Delta t_{\mathrm{i}})/\sigma_{\mathrm{TOF, i}}$ distribution, where $\Delta t$ is the measured time-of-flight, $\Delta t_{\mathrm{i}}$ is the expected value for the different species, and $\sigma_{\mathrm{TOF, i}}$ is the detector resolution for the considered species. The distribution is fitted with an exponentially modified Gaussian function with a tail on the right, while the background derived from mismatched tracks is parameterized with a single or double exponential (depending on the \pt interval) for the (anti)deuterons or with a second-order polynomial for the antiprotons. The signal extraction is performed by integrating the distribution in the interval $[\mu_0-3\sigma_{\mathrm{TOF}}, \mu_0+3.5\sigma_{\mathrm{TOF}}]$, where $\mu_0$ is the mean of the Gaussian signal, which corrects for small miscalibration effects in the \TOF that have been observed in a previous analyses~\cite{ALICE_nuclei_antinuclei, ALICE_nuclei_pPb_8TeV, ALICE_pp_B2_jet}. The asymmetric integration interval is chosen to take into account the exponential tail of the distribution, mainly due to tracks inducing signals in more than one detector readout element.

\subsection{Corrections}
\label{subsec:corrections}

The obtained raw distributions are corrected for the reconstruction efficiency and the secondary-particle contribution using MC simulated events. In particular, for the reconstruction efficiency, a MC simulation with light (anti)nuclei generated with a flat \pt distribution in the transverse-momentum range $0 < \pt < 10$~\GeVc and a flat rapidity distribution in the range $-1< \ycms < 1$ is used. The generated particles are injected on top of the underlying event generated by HIJING~\cite{HIJING}. The interactions of the generated particles with the experimental apparatus are modeled with the Geant4 package~\cite{GEANT4}. For the secondary particles evaluation, instead, a MC simulation where the particle production is described by DPMJET~\cite{DPMJET} is used.  

\subsubsection{Efficiency and acceptance}
\label{subsubsec:efficiency}
The Acceptance $\times$ Efficiency ($A\times\epsilon$) correction is calculated as the ratio of reconstructed and generated (anti)nuclei momentum distributions in the simulation, with the same track selections as used for the data applied in the reconstructed MC. The obtained efficiency for (anti)deuterons ranges from 45 to 80\% in the \pt interval $0.6 < \pt < 1.2$~\GeVc, where only the information from the \TPC is used. It is about 50\% in the remaining transverse-momentum range, where the particle identification is complemented with the \TOF information. Moreover, in all transverse-momentum ranges, the efficiency for antideuterons is smaller than that for deuterons, due to the annihilation processes with the detector material. For antiprotons the obtained values are between 25 and 65\% for transverse momentum below 0.7 \GeVc, while for $0.7 < \pt < 1.5$~\GeVc the efficiency ranges from 30 to 40\%. 

\subsubsection{Secondary particles from spallation and weak decays}
\label{subsubsec:secondary}
Primary particles can interact with the detector material and produce nuclear fragments, called spallation products. Since antinuclei cannot originate from the detector material, this contamination only applies to deuterons. The fraction of secondary nuclei from the material is obtained from template fits to the measured $\mathrm{DCA}_{xy}$ distributions, as done in previous measurements~\cite{ALICE_nuclei_multiplicity_pPb_5TeV, ALICE_deuteron_pp_13TeV} and described in Ref.~\cite{ALICE_nuclei_pp_PbPb}. Antideuterons from data are used as a template for primary deuterons, while the template for secondary deuterons from spallation is taken from MC simulations.  
The fit is executed in the range $\lvert\mathrm{DCA}_{xy}\rvert <  1.0$~cm. The obtained fraction of secondary deuterons varies from 45\% in the lowest \pt range, and approaches zero for $\pt > 1.8$~\GeVc. 

For antiprotons the secondary contamination arises from weak decays of strange particles, such as $\Lambda$ and $\Sigma$ baryons. A procedure similar to that used for deuterons is applied to antiprotons, although in this case both the primary and secondary templates are obtained using MC simulations. The fit is performed in the same $\mathrm{DCA}_{xy}$ intervals, and the secondary fraction, of about 20\%, is found to be weakly dependent on \pt. 

\subsection{Systematic uncertainties}
\label{subsec:syst}
The different contributions to the systematic uncertainties of (anti)deuterons and antiprotons are reported in Tab.~\ref{tab:errors}. The final value is obtained as the sum in quadrature of all contributions, and it varies from 3 to 5\% for (anti)deuterons and is about 3\% for antiprotons. 
\begin{table}
\centering
\caption{Summary of the different contributions of the systematic uncertainties, both for (anti)deuterons and antiprotons, in the lowest and highest \pt intervals. All values are given as percentages.}
\label{tab:errors}
\begin{tabular}{cccccc}
\toprule
Particle & &\multicolumn{2}{c}{d ($\bar{\mathrm{d}}$)} & \multicolumn{2}{c}{$\bar{\mathrm{p}}$}\\
\pt range (\GeVc) & & 0.6 -- 0.8 & 2.4 -- 3.0 & 0.3 -- 0.4 & 1.4 -- 1.5\\
\midrule 
Source of uncertainty & & & & & \\
\midrule
Tracking & &0.3 (1.5) & 0.5 (0.9) & 0.7 & 2.4 \\
ITS-TPC matching & & 1.0 (1.0) & 1.0 (1.0) & 1.0 & 1.0 \\
Material budget & & 1.0 (1.0) & 1.0 (1.0) & 1.0 & 1.0 \\
Hadronic interaction & & 0.4 (3.0) & 0.7 (4.0) & 0.3 & 1.4 \\
Primary fraction & & 5.0 (-) & negl. (-) & 1.0 & 1.1 \\
\multirow{3}{*}{	Signal extraction} & Toward & 1.4 (1.0) & 1.8 (2.6) & 0.7 & 0.5  \\
								   & Transverse & 1.4 (1.9) & 3.7 (2.8) & 0.6 & 0.5  \\
								   & Away & 2.1 (1.5) & 2.8 (2.1) & 0.7 & 0.6 \\
								   
\bottomrule
\end{tabular}
\end{table}

For the systematic uncertainty of the track reconstruction, the quality criteria used in the analysis and reported in Sect.~\ref{subsec:pid} are varied 50 times both in data and MC simulations using a random uniform distribution around each nominal value. The width of the variation differs for the individual quality criteria as sources of uncertainty, while maintaining a good overall quality of the reconstructed tracks. The relative systematic uncertainty is given by the standard deviation over the mean value of the distribution of the selected particles, considering the corrected yield for each transverse-momentum interval. This contribution is between 0.3\% (1.5\%) and 0.5\% (0.9\%) for deuterons (antideuterons), while it varies from 0.7 to 2.4\% for antiprotons. At low \pt, the uncertainty for antideuterons is larger than for deuterons, due to statistical fluctuations in the applied procedure.

The uncertainty due to the ITS-TPC matching efficiency is evaluated as the difference between data and MC. The obtained value, of about 1\% for both species, is \pt independent.

The systematic uncertainty due to material budget, equal to 1\% for both (anti)deuterons and antiprotons, is taken from Ref.~\cite{ALICE_deuteron_pp_13TeV}. This source is estimated by comparing different MC simulations in which the material budget is increased and decreased by 4.5\%, which is the uncertainty in determining the material budget of the detector~\cite{ALICE_description_2}. 

The uncertainty due to the hadronic interaction cross section with the detector material is estimated by varying the default inelastic cross section of the selected species in Geant4 within its uncertainty, obtained with measurements on different targets~\cite{hadronic_1, hadronic_2, hadronic_3, hadronic_4, hadronic_5, hadronic_6, hadronic_7, hadronic_8, hadronic_9}. The obtained value is \pt dependent, since at high transverse momentum the particle identification is obtained using different detectors, hence different hadronic processes may contribute. Moreover, matter and antimatter are subjected to different hadronic interactions with the detector material, with different cross section values. For this reason, this systematic uncertainty for deuteron and antideuteron has different values. The obtained uncertainty varies from 0.4\% (3.0\%) to 0.7\% (4.0\%) for deuterons (antideuterons), while for antiprotons it rises from 0.3\% at low \pt to 1.4\% at high \pt. 

The systematic uncertainty on the primary fraction is obtained by varying the longitudinal and transverse DCA selection criteria, the intervals width, and the fit interval for the template method. The obtained value for each case is equal to the difference between the highest and lowest values divided by two. This source of systematic uncertainty is negligible for antideuterons (due to the fact that antinuclei are not produced by spallation processes), while it is relevant for deuterons at low transverse momentum, where it reaches 5\%. The same contribution for the antiproton is weakly dependent on \pt and equals about 1\%.

Finally, the systematic uncertainty due to the signal extraction, calculated for the three azimuthal regions due to the azimuthal dependence of the signal extraction procedure in the analysis, is evaluated with different methods. These include varying the integration interval (considering 50 variations of the integration interval where the extremes are uniformly generated in the range 
[$-3.5n\sigma_{\TPC}, -2.0n\sigma_{\TPC}$] and [$2.0n\sigma_{\TPC}, 3.5n\sigma_{\TPC}$] for the TPC, and [$-3.0n\sigma_{\TOF}, -2.0n\sigma_{\TOF}$] and [$3.0n\sigma_{\TOF}, 4.5n\sigma_{\TOF}$] for the \TOF), using a different parametrization for the background functions (like Gaussian or exponential function for the TPC), and varying the signal extraction strategy (considering bin counting or the integral of the signal function). For the former, the obtained value is equal to the ratio between the standard deviation and the mean of the distribution of the corrected yield, where the efficiency is also calculated considering the different integration intervals. For the others, half of the difference  between the obtained value and the nominal value is considered. This source is different in the three azimuthal regions, ranging from a minimum of 1.4\% to a maximum of 3.7\% for (anti)deuterons, while for antiprotons it ranges between 0.5\% and 0.7\%. 

\section{Results}
\label{sec:result}

\begin{figure}
\centering
\subfigure{\includegraphics[width=0.49\textwidth]{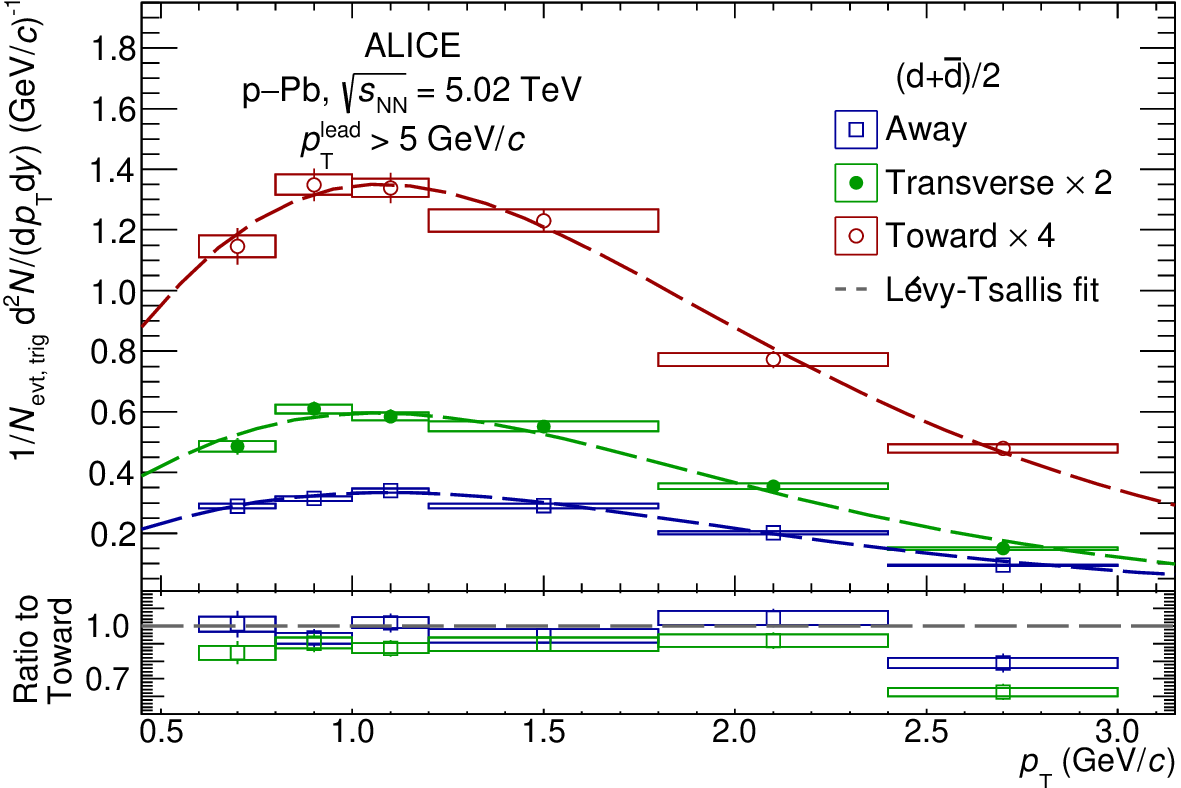}}
\subfigure{\includegraphics[width=0.49\textwidth]{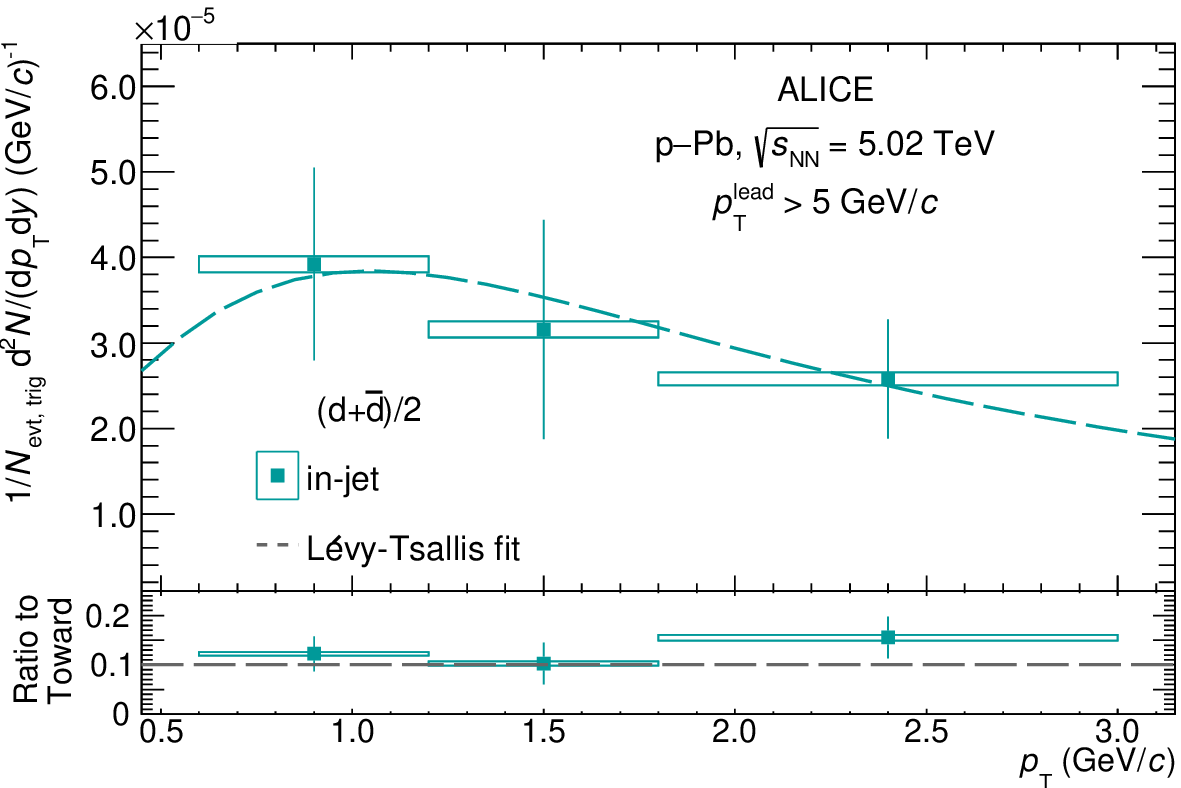}}
\subfigure{\includegraphics[width=0.49\textwidth]{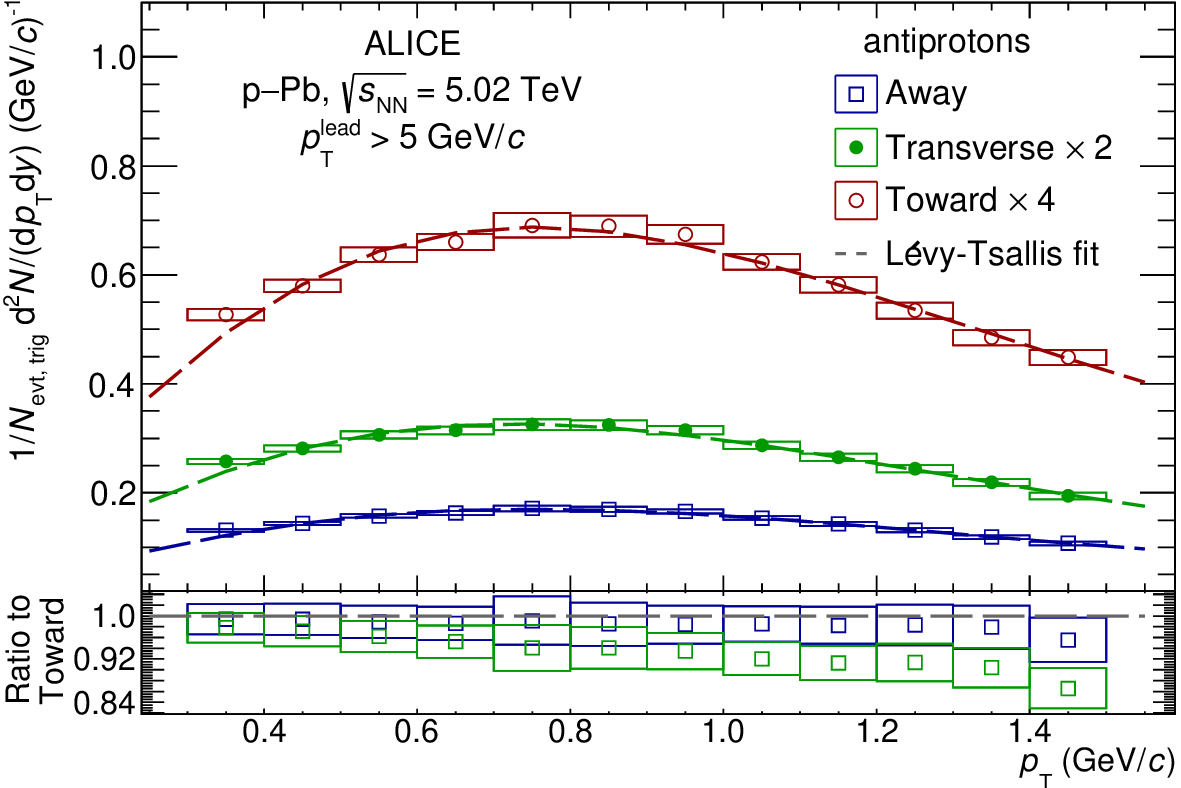}}
\subfigure{\includegraphics[width=0.49\textwidth]{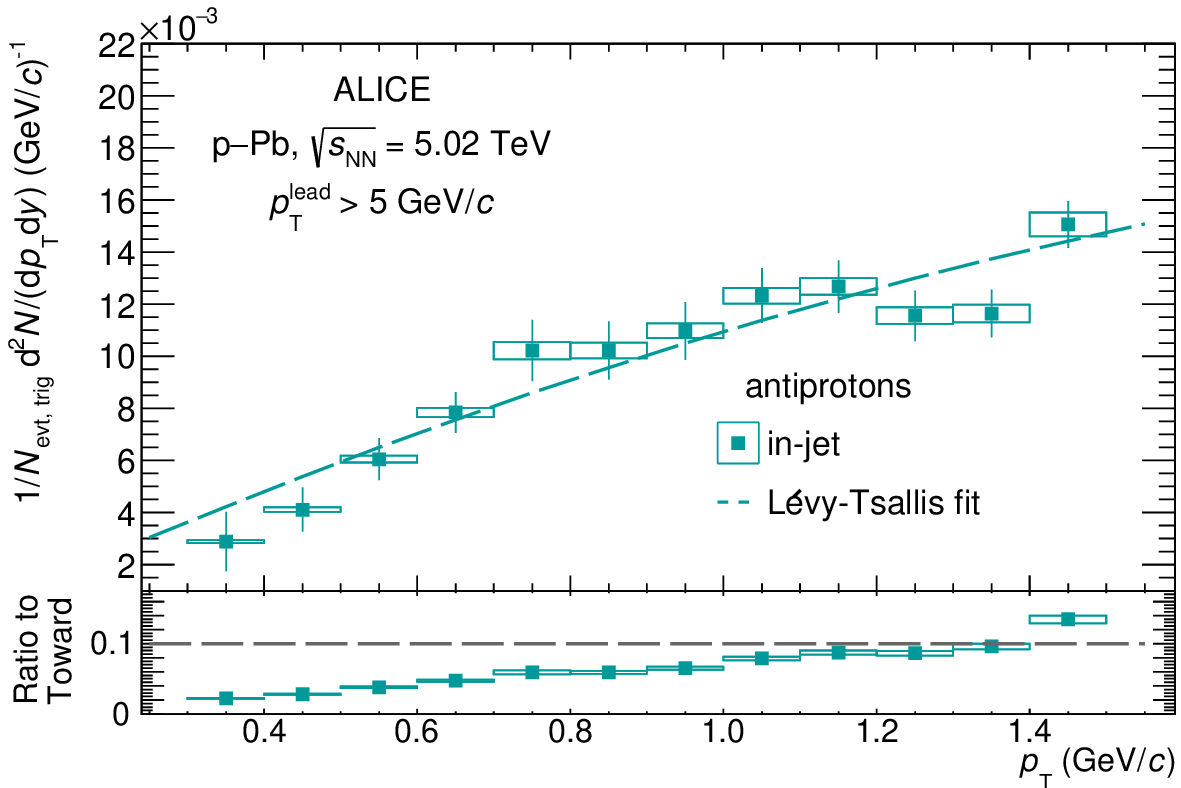}}
\caption{The average of deuteron and antideuteron (top row) and antiproton (bottom row) \pt -differential yields in the three azimuthal regions (left) and in jets (right) measured in \pPb collisions at \fivenn. Statistical and systematical uncertainties are reported, respectively, with vertical lines and boxes. Individual \LT fits are also shown. The distributions are scaled for visibility reasons. The bottom panel of each figure shows the ratio of the distribution with respect to the Toward one.}
\label{fig:spectra}
\end{figure}

The \pt distributions of the average of deuteron and antideuteron in the three azimuthal regions are shown in Fig.~\ref{fig:spectra} (top-left) while the distributions for antiprotons are displayed in the bottom left panel. The dotted lines represent \LT fits~\cite{Levy_Tsallis}. For the mean of deuteron and antideuteron, the total systematic uncertainty is propagated considering the two errors to be independent. The distributions in the different regions are scaled for visibility reasons. For both species, the transverse-momentum distributions in the Toward and Away regions are compatible, while the Transverse is about 90\% of the Toward region. These experimental results are consistent with the one already observed in \pp collisions~\cite{ALICE_pp_B2_jet}.
The in-jet transverse-momentum distribution is obtained by subtracting the distribution in the Transverse region (which is dominated by the underlying event) from the distribution in the Toward region (which contains both jet fragmentation and underlying event). These distributions for both the average between deuteron and antideuteron and antiproton are reported in  Fig.~\ref{fig:spectra} (right). In the case of the mean of the deuteron and antideuteron, due to the reduced data sample available, the TPC transverse-momentum intervals and the two last TOF \pt intervals have been merged using a weighted mean, using as weight the value of the \LT function. The total systematic uncertainty for the mean of deuteron and antideuteron in-jet spectrum is calculated considering the mean of the sum in quadrature of the systematic uncertainty of the two particle species. For each species, the total in-jet systematic uncertainty is equal to the sum in quadrature of the contribution listed in Tab.~\ref{tab:errors}, where the signal extraction one is calculated as the sum in quadrature of the Toward and Transverse regions. A similar approach has been applied to the antiproton in-jet spectrum. As already observed in \pp collisions~\cite{ALICE_pp_B2_jet, ALICE_deuteron_jets}, the jet contribution accounts only for about 10\% of the total particle production in the Toward region. Hence, with the jet reconstruction adopted in this analysis, also the Toward region is dominated by the underlying event. Moreover, as expected in the case of the jet-associated particle production~\cite{ALICE_jet_charged_particle_production}, the distributions in jets show a hardening with respect to the one out of jets. It is worth to notice that these distributions can be influenced by the long range correlations and the different elliptic flow of the (anti)deuterons and antiprotons. 

\begin{figure}
\centering
\includegraphics[width=0.6\textwidth]{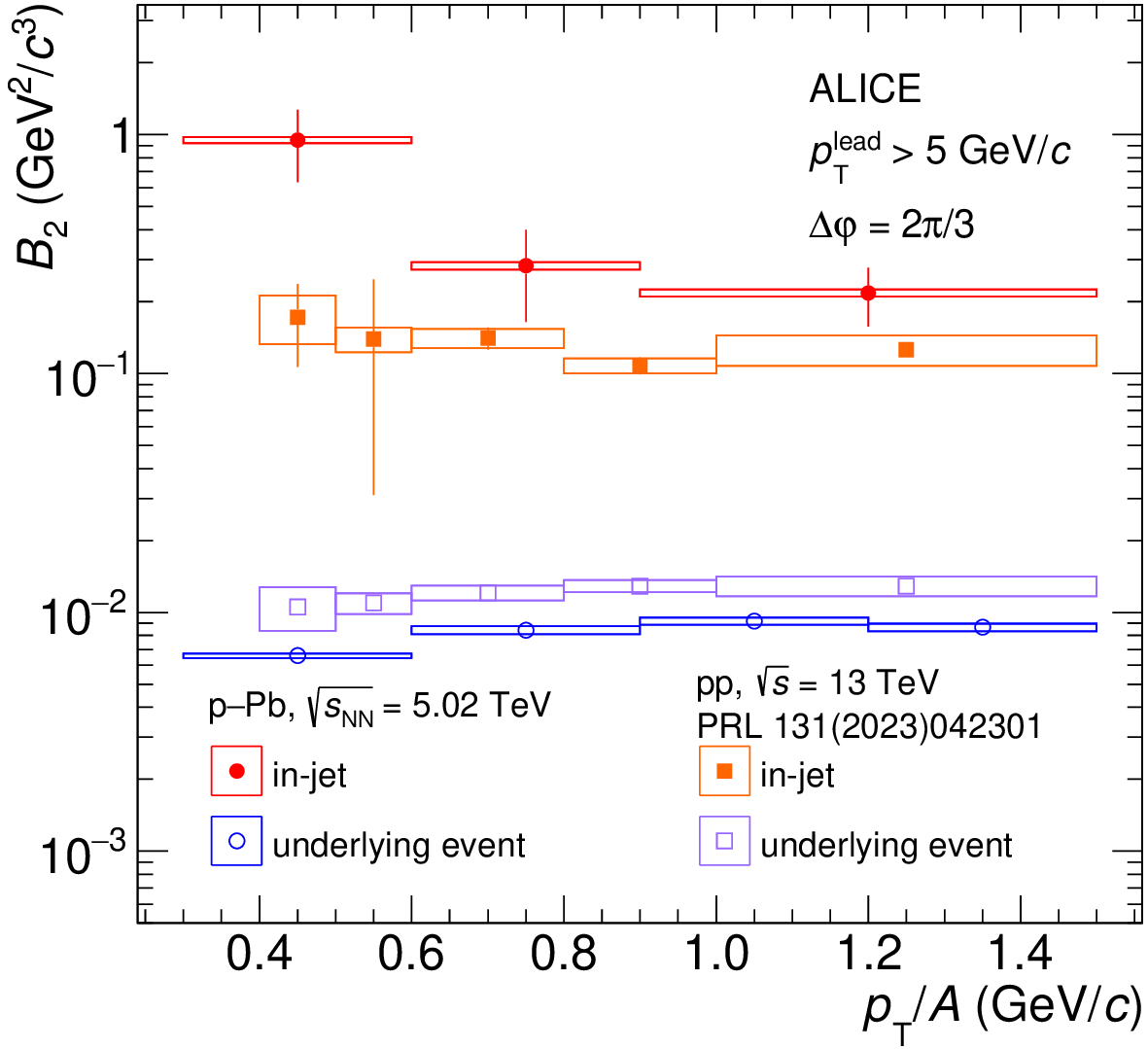}
\caption{The coalescence parameter in jets \coaljet and in the underlying event \coalUE as a function of the per nucleon transverse momentum \pt/$A$ (here, $A = 2$) in \pp collisions at \thirtheen, taken from~\cite{ALICE_pp_B2_jet}, and \pPb collisions at \fivenn. Statistical and systematic uncertainties are represented, respectively, by vertical bars and boxes.}
\label{fig:coalescence}
\end{figure}

With the obtained distributions, the coalescence parameter in jets (\coaljet) and in the underlying event (\coalUE) is calculated according to Eq.~\ref{eqn:coalescence_parameter}. This quantity is shown in Fig.~\ref{fig:coalescence} as a function of the per-nucleon transverse momentum, \pt/$A$, where $A = 2$ is the mass number of the deuteron. The figure shows both \coaljet (full markers) and \coalUE (empty markers) in \pp collisions at \thirtheen (square markers), taken from~\cite{ALICE_pp_B2_jet}, and in \pPb collisions at \fivenn (circle markers).
The coalescence parameter \coaljet is enhanced with respect to \coalUE in \pPb collisions by more than a factor of 20 with a significance of about $4.6\sigma$, larger than the enhancement of about 15, observed in pp collisions~\cite{ALICE_pp_B2_jet}. One explanation provided by the coalescence model for this enhancement is the reduced distance in phase space between nucleons in jets compared with the underlying event. 

More information can be obtained if the values observed in the two collision systems are compared in the single azimuthal regions. In particular, in the underlying event region, the \coalUE in \pPb collisions is lower with respect to the same quantity in \pp collisions. This can be explained by considering the dependence of the coalescence parameter on the source size. According to the coalescence model~\cite{coal_3}, the coalescence parameter decreases with increasing source size, since the nucleons are naturally more spread out in phase space. The source size of the \pp and \pPb colliding systems, in particular the values of the source radii, have been measured by ALICE through femtoscopy techniques~\cite{ALICE_femto}. The obtained values are of about 1~fm in \pp collisions~\cite{ALICE_source_pp} and 1.5~fm in \pPb collisions~\cite{ALICE_source_pPb}. 

The different values of the coalescence parameter in jets, where the \pPb values are larger than in \pp collisions, do not have a unique explanation at the moment. They could arise from different particle-species composition of jets in the two systems, stronger momentum correlation between the nucleons, or a higher average \pt value in jets. Moreover, a dependence on \pt/$A$ is observed. In order to distinguish among these different explanations, specific analyses that investigate these hypotheses and a larger data sample might be considered, since the in-jet results are dominated by the statistical uncertainty. 

\begin{figure}
\centering
\includegraphics[width=0.65\textwidth]{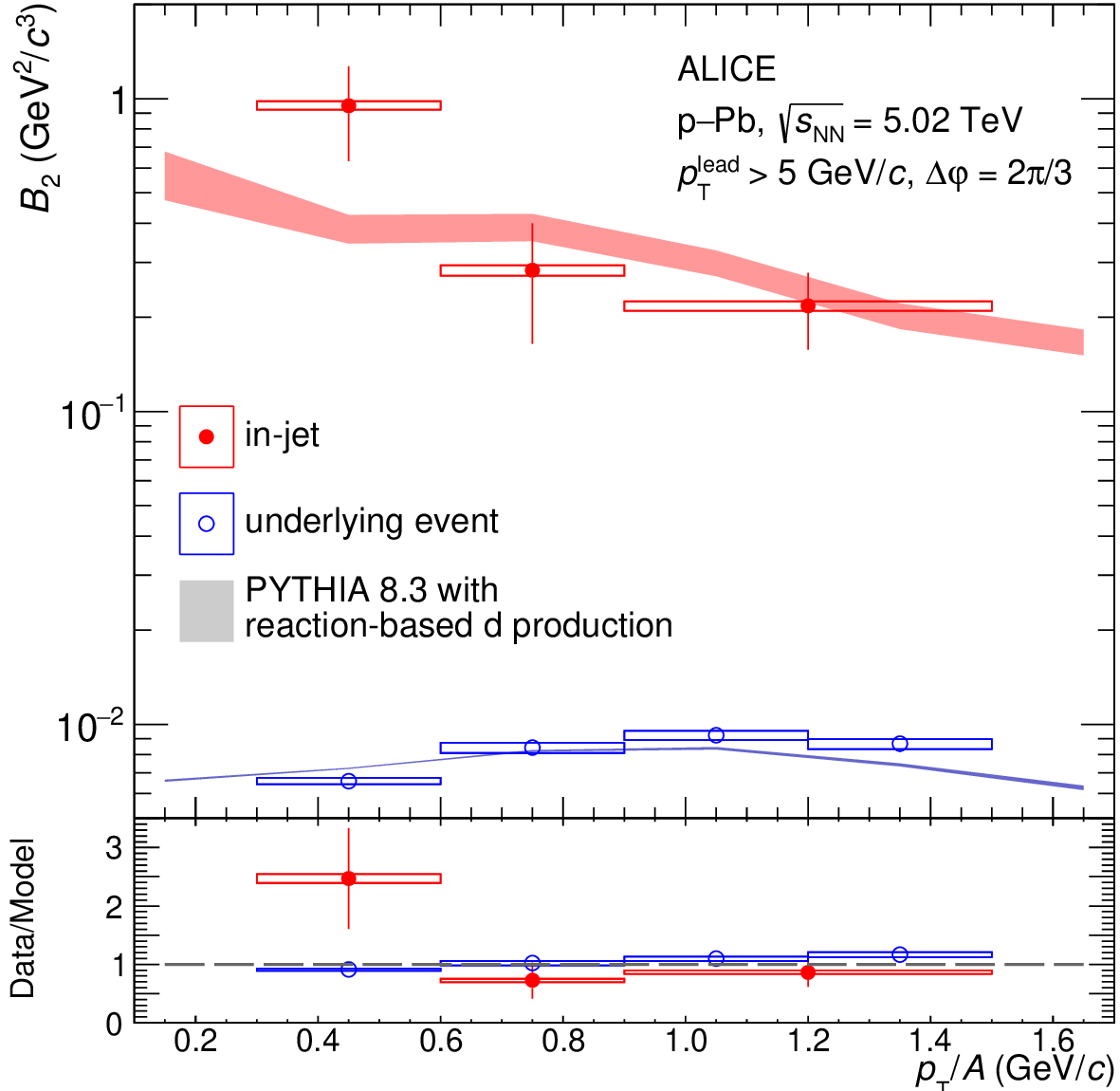}
\caption{The coalescence parameter in jets \coaljet and in the underlying event \coalUE in \pPb collisions at \fivenn as a function of \pt/$A$ compared with the model prediction given by PYTHIA 8.314 with Angantyr.}
\label{fig:models}
\end{figure}

The obtained results are compared with the predictions from a reaction-based deuteron production model implemented in PYTHIA 8.314~\cite{PYTHIA} with the Angantyr model~\cite{Angantyr}. Angantyr, inspired by the Fritiof model~\cite{Fritiof_1, Fritiof_2}, uses the Glauber model to calculate the number of wounded nucleons and generate the hadronic final state in proton-nucleus and nucleus-nucleus collisions. Deuteron production is implemented in PYTHIA 8.314 with a nuclear reactions-based model reported in detail in Refs.~\cite{PYTHIA, ALICE_pp_B2_jet}. Only statistical uncertainties are computed in the model. As shown in Fig.~\ref{fig:models}, the model qualitatively reproduces the data, as already observed in pp collisions~\cite{ALICE_pp_B2_jet}. In particular, the model is able to reproduce the enhancement of \coaljet with respect to \coalUE. While the predictions for \coalUE are consistent with the experimental data, the \coaljet seems to be underestimated by the model at low \pt/$A$ values, though the current statistical uncertainties do not allow a conclusive statement. The comparison with such a model, even if a state-of-the-art coalescence approach is not implemented, gives significant insights in the deuteron production, considering the recent results reported in Ref.~\cite{ALICE_deuteron_femto}. In such study, in fact, the deuteron-pion correlation function shows that the deuteron is produced through nuclear fusion processes, and PYTHIA 8.314 uses a similar approach for the deuteron production. 

\begin{figure}
\centering
\includegraphics[width=0.8\textwidth]{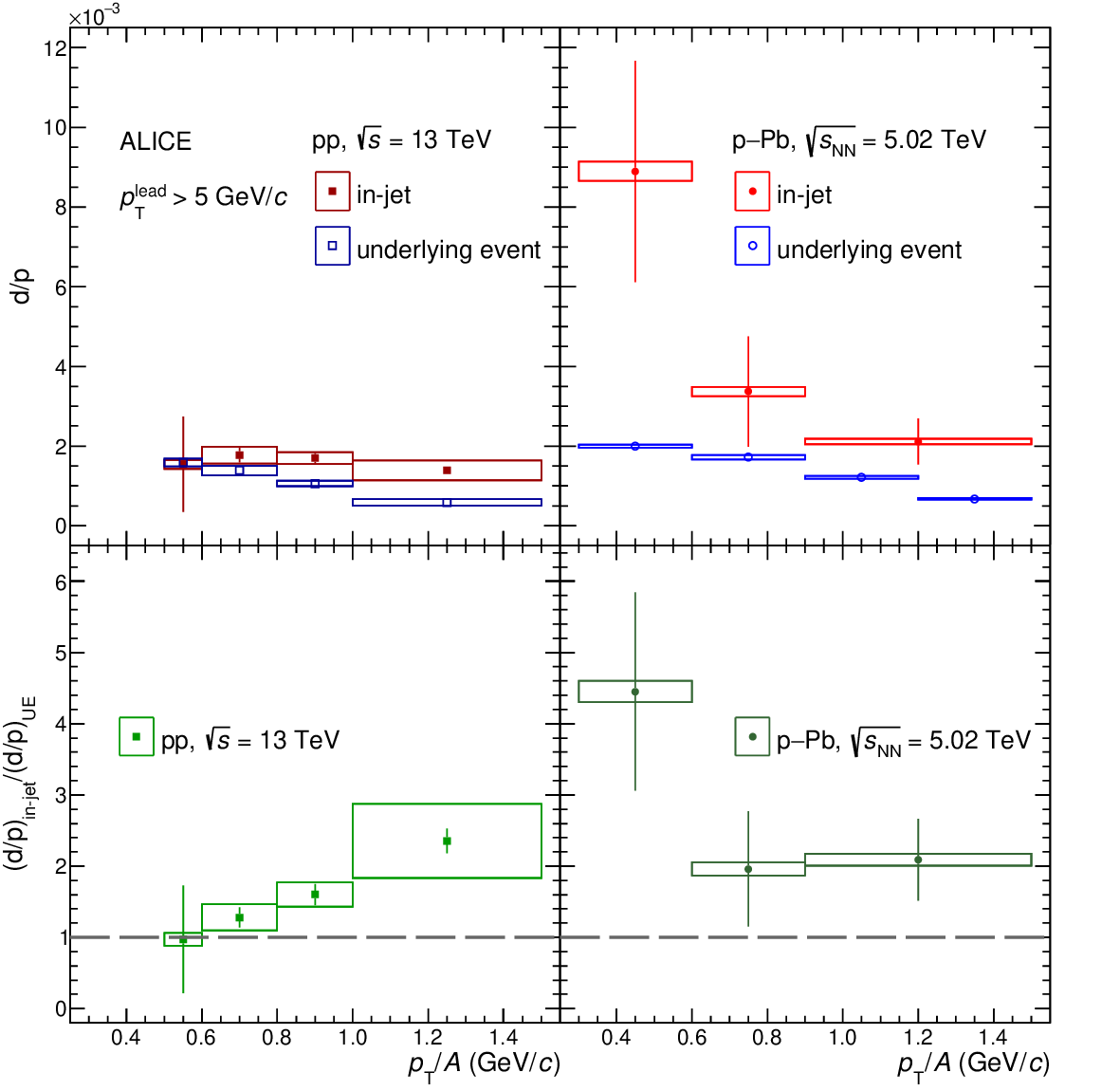}
\caption{Top: The deuteron-to-proton yield ratio in \pp collisions at \thirtheen~\cite{ALICE_pp_B2_jet, ALICE_proton_RT} (left) and \pPb collisions at \fivenn (right) in jets and in the underlying event. Bottom: The double ratio $(\mathrm{d/p})_{\mathrm{in-jet}}/(\mathrm{d/p})_{\mathrm{UE}}$ in \pp collisions at \thirtheen (left) and \pPb collisions at \fivenn (right). Statistical and systematic uncertainties are represented, respectively, by vertical bars and boxes.}
\label{fig:ratios_systems}
\end{figure}

The ratio of the measured yields of deuterons and protons is also sensitive to the production mechanism of light nuclei. In particular, this ratio can give more insights in hadrochemistry, exploring the differences across different collision systems. The deuteron-to-proton yield ratio as a function of the reduced transverse momentum \pt/$A$ is reported in the top row of Fig.~\ref{fig:ratios_systems} for both \pp collisions at \thirtheen (left panels) and \pPb collisions at \fivenn (right panels)  both in the jet region (red, full marker) and in the underlying event (blue, open marker). 
The deuteron and proton distributions in the \pp case are taken, respectively, from Refs.~\cite{ALICE_pp_B2_jet, ALICE_proton_RT}. In the ratio, the statistical uncertainties are treated as independent. In both collision systems, the ratio of deuterons to protons is higher in jets with respect to the underlying event, as reported in the bottom row in Fig.~\ref{fig:ratios_systems}, by a factor between one and two in \pp collisions and between two and four in \pPb collisions. These results suggest that the deuteron production is favored in jets with respect to the production out of jets, which could be explained by considering the enhanced coalescence probability in jets.

\begin{figure}
    \centering
    \includegraphics[width=0.8\textwidth]{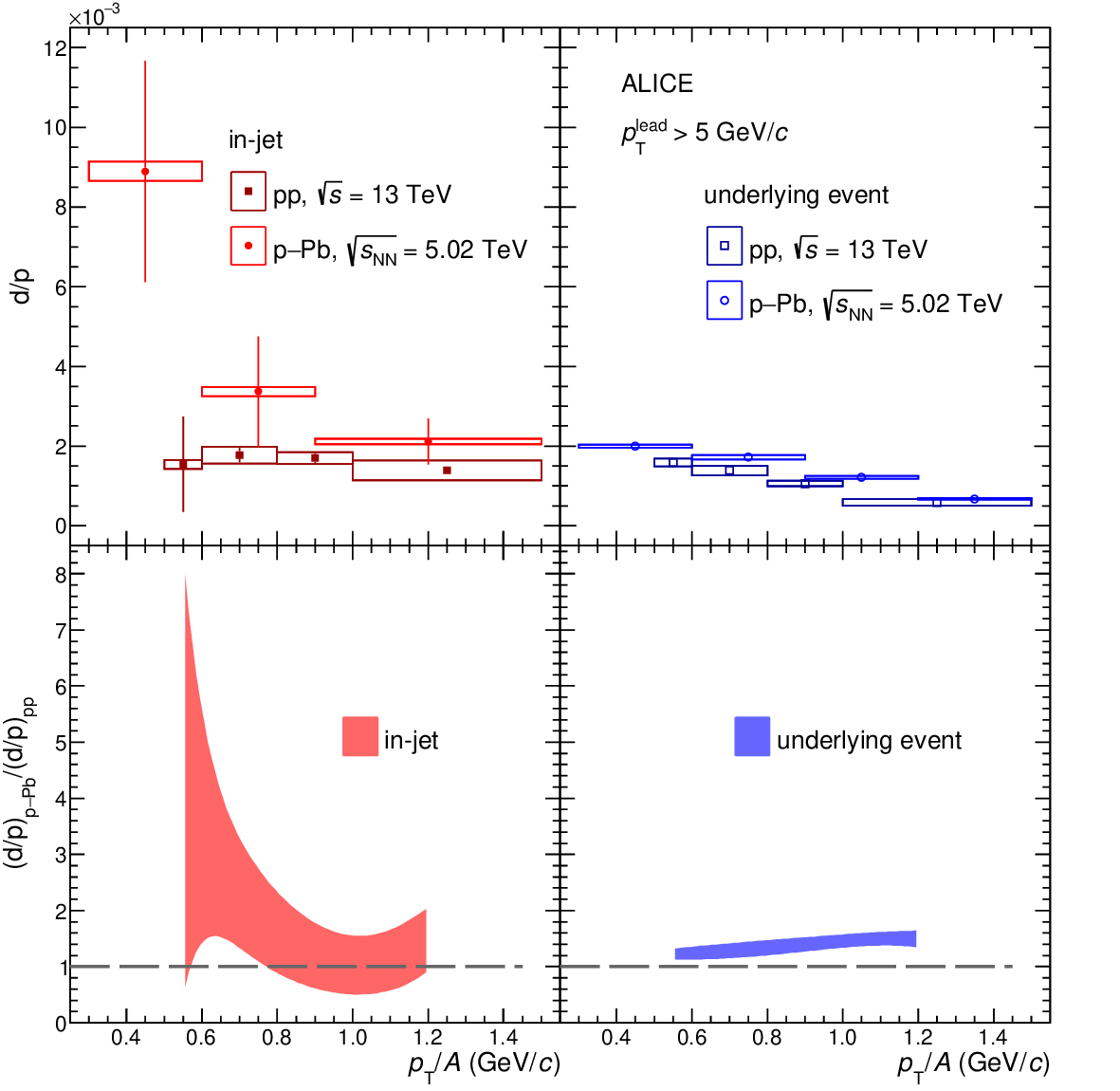}
    \caption{Top: The deuteron-to-proton yield ratio in jets (left) and in the underlying event (right) in \pp collisions at \thirtheen~\cite{ALICE_pp_B2_jet, ALICE_proton_RT} and \pPb collisions at \fivenn (right). Statistical and systematic uncertainties are represented, respectively, by vertical bars and boxes. Bottom: The double ratio $(\mathrm{d/p})_{\pPb}/(\mathrm{d/p})_{\mathrm{pp}}$ in jets (left) and in the  underlying event (right). The band represents the statistical and systematic uncertainties summed in quadrature.}
    \label{fig:ratios_regions}
\end{figure}

A further comparison is made by considering the ratio of the deuteron to proton ratios of the two collision systems in the different azimuthal regions. This comparison is reported in Fig.~\ref{fig:ratios_regions}.
In this case, since the \pt intervals of the two analyses are not the same, an interpolation of the experimental points is used to evaluate the ratio. The systematic uncertainties are treated as independent when propagating them through the double ratio. Finally, the uncertainty of the double ratio is equal to the quadrature sum of the statistical and systematic uncertainties. For the in-jet region the statistical uncertainty dominates, being about 90\% of the total error. For the underlying event region, in contrast, statistical and systematic uncertainties are comparable in each interval of \pt/$A$. The results, reported in the bottom row of Fig.~\ref{fig:ratios_regions}, show that in both jets and the underlying event the double ratio is compatible with unity within the 1$\sigma$ uncertainty, and therefore no firm conclusion on the jet hadrochemistry in different systems can be drawn. To fully address the observed results further detailed studies of jet hadrochemistry are required, as the present results do not allow a definitive statement due to the limited jet-spectra data sample and the dominance of statistical uncertainties in the analysis.

\section{Summary and conclusions}
This article presents the measurement of (anti)deuteron and antiproton distributions, and the deuteron coalescence parameter in and out of jets in \pPb collisions at \fivenn, complementing and extending the study presented in Ref.~\cite{ALICE_pp_B2_jet} for \pp collisions. This experimental observation is consistent with the coalescence picture that predicts an enhanced formation probability of a bound state for nucleons that have a small average phase-space distance. The results are compared with previous measurements from \pp collisions at \thirtheen. The enhancement of \coaljet with respect to \coalUE measured in \pPb collisions, a factor more than 20 with a $4.6\sigma$ significance, is larger with respect to the one observed in \pp collisions, of about 15. Such a result could be explained by the different source sizes and, possibly, by different particle-species compositions of jets in the two collision systems. Moreover, the deuteron-to-proton ratio is enhanced in jets with respect to the underlying event in both \pp and \pPb collisions by a factor between one and two in the \pp case and between two and four in \pPb collisions. Comparing the d/p values in jets and in the underlying event between the two different collision systems, their ratio is compatible with unity within the 1$\sigma$ uncertainty, hence no firm conclusion on the jet hadrochemistry in different systems can be drawn. The obtained results on the coalescence parameter are compared with the predictions from PYTHIA 8.314 with the Angantyr model, which is able to qualitatively reproduce the large gap in the observable, even if \coaljet seems to be underestimated by the model at low \pt/$A$. However, the current statistical uncertainties do not allow for a conclusive statement.

Future studies exploiting the larger data sample from LHC Run 3 will be extended to heavier (anti)nuclei, such as (anti)$^3\mathrm{He}$, applying a multi-differential approach by varying the $p_{\mathrm{T}}^{\mathrm{lead}}$ selection, considering different transverse particle multiplicity, and  fully reconstructing jets with a jet-finder algorithm. Further investigations of the coalescence parameter in and out of jets will provide additional information to constrain the light (anti)nuclei formation process, which could be used as input for further experimental measurements of the production of nuclei in colliders and in space-based experiments. 


\newenvironment{acknowledgement}{\relax}{\relax}
\begin{acknowledgement}
\section*{Acknowledgements}
\input{fa_2026-01-22_Opt_C.tex}
\end{acknowledgement}

\bibliographystyle{utphys}   
\bibliography{bibliography}

\newpage
\appendix

%
%

\section{The ALICE Collaboration}
\label{app:collab}
\input{Alice_Authorlist_2026-01-22_Opt_C.tex}  
\end{document}

%% file: commands.tex
%

\newcommand{\pp}           {pp\xspace}
\newcommand{\ppbar}        {\mbox{$\mathrm {p\overline{p}}$}\xspace}
\newcommand{\XeXe}         {\mbox{Xe--Xe}\xspace}
\newcommand{\PbPb}         {\mbox{Pb--Pb}\xspace}
\newcommand{\pA}           {\mbox{pA}\xspace}
\newcommand{\pPb}          {\mbox{p--Pb}\xspace}
\newcommand{\AuAu}         {\mbox{Au--Au}\xspace}
\newcommand{\dAu}          {\mbox{d--Au}\xspace}

\newcommand{\s}            {\ensuremath{\sqrt{s}}\xspace}
\newcommand{\snn}          {\ensuremath{\sqrt{s_{\mathrm{NN}}}}\xspace}
\newcommand{\pt}           {\ensuremath{p_{\rm T}}\xspace}
\newcommand{\meanpt}       {$\langle p_{\mathrm{T}}\rangle$\xspace}
\newcommand{\ycms}         {\ensuremath{y_{\rm CMS}}\xspace}
\newcommand{\ylab}         {\ensuremath{y_{\rm lab}}\xspace}
\newcommand{\etarange}[1]  {\mbox{$\left | \eta \right |~<~#1$}}
\newcommand{\yrange}[1]    {\mbox{$\left | y \right |~<~#1$}}
\newcommand{\dndy}         {\ensuremath{\mathrm{d}N_\mathrm{ch}/\mathrm{d}y}\xspace}
\newcommand{\dndeta}       {\ensuremath{\mathrm{d}N_\mathrm{ch}/\mathrm{d}\eta}\xspace}
\newcommand{\avdndeta}     {\ensuremath{\langle\dndeta\rangle}\xspace}
\newcommand{\dNdy}         {\ensuremath{\mathrm{d}N_\mathrm{ch}/\mathrm{d}y}\xspace}
\newcommand{\Npart}        {\ensuremath{N_\mathrm{part}}\xspace}
\newcommand{\Ncoll}        {\ensuremath{N_\mathrm{coll}}\xspace}
\newcommand{\dEdx}         {\ensuremath{\textrm{d}E/\textrm{d}x}\xspace}
\newcommand{\RpPb}         {\ensuremath{R_{\rm pPb}}\xspace}

\newcommand{\nineH}        {$\sqrt{s}~=~0.9$~Te\kern-.1emV\xspace}
\newcommand{\seven}        {$\sqrt{s}~=~7$~Te\kern-.1emV\xspace}
\newcommand{\twoH}         {$\sqrt{s}~=~0.2$~Te\kern-.1emV\xspace}
\newcommand{\twosevensix}  {$\sqrt{s}~=~2.76$~Te\kern-.1emV\xspace}
\newcommand{\five}         {$\sqrt{s}~=~5.02$~Te\kern-.1emV\xspace}
\newcommand{\thirtheen}    {$\sqrt{s}~=~13$~Te\kern-.1emV\xspace}
\newcommand{\twosevensixnn}{$\sqrt{s_{\mathrm{NN}}}~=~2.76$~Te\kern-.1emV\xspace}
\newcommand{\fivenn}       {$\sqrt{s_{\mathrm{NN}}}~=~5.02$~Te\kern-.1emV\xspace}
\newcommand{\LT}           {L{\'e}vy-Tsallis\xspace}
\newcommand{\GeVc}         {Ge\kern-.1emV/$c$\xspace}
\newcommand{\MeVc}         {Me\kern-.1emV/$c$\xspace}
\newcommand{\TeV}          {Te\kern-.1emV\xspace}
\newcommand{\GeV}          {Ge\kern-.1emV\xspace}
\newcommand{\MeV}          {Me\kern-.1emV\xspace}
\newcommand{\GeVmass}      {Ge\kern-.2emV/$c^2$\xspace}
\newcommand{\MeVmass}      {Me\kern-.2emV/$c^2$\xspace}
\newcommand{\lumi}         {\ensuremath{\mathcal{L}}\xspace}

\newcommand{\ITS}          {\rm{ITS}\xspace}
\newcommand{\TOF}          {\rm{TOF}\xspace}
\newcommand{\ZDC}          {\rm{ZDC}\xspace}
\newcommand{\ZDCs}         {\rm{ZDCs}\xspace}
\newcommand{\ZNA}          {\rm{ZNA}\xspace}
\newcommand{\ZNC}          {\rm{ZNC}\xspace}
\newcommand{\SPD}          {\rm{SPD}\xspace}
\newcommand{\SDD}          {\rm{SDD}\xspace}
\newcommand{\SSD}          {\rm{SSD}\xspace}
\newcommand{\TPC}          {\rm{TPC}\xspace}
\newcommand{\TRD}          {\rm{TRD}\xspace}
\newcommand{\VZERO}        {\rm{V0}\xspace}
\newcommand{\VZEROA}       {\rm{V0A}\xspace}
\newcommand{\VZEROC}       {\rm{V0C}\xspace}
\newcommand{\Vdecay} 	   {\ensuremath{V^{0}}\xspace}

\newcommand{\ee}           {\ensuremath{e^{+}e^{-}}} 
\newcommand{\pip}          {\ensuremath{\pi^{+}}\xspace}
\newcommand{\pim}          {\ensuremath{\pi^{-}}\xspace}
\newcommand{\kap}          {\ensuremath{\rm{K}^{+}}\xspace}
\newcommand{\kam}          {\ensuremath{\rm{K}^{-}}\xspace}
\newcommand{\pbar}         {\ensuremath{\rm\overline{p}}\xspace}
\newcommand{\kzero}        {\ensuremath{{\rm K}^{0}_{\rm{S}}}\xspace}
\newcommand{\lmb}          {\ensuremath{\Lambda}\xspace}
\newcommand{\almb}         {\ensuremath{\overline{\Lambda}}\xspace}
\newcommand{\Om}           {\ensuremath{\Omega^-}\xspace}
\newcommand{\Mo}           {\ensuremath{\overline{\Omega}^+}\xspace}
\newcommand{\X}            {\ensuremath{\Xi^-}\xspace}
\newcommand{\Ix}           {\ensuremath{\overline{\Xi}^+}\xspace}
\newcommand{\Xis}          {\ensuremath{\Xi^{\pm}}\xspace}
\newcommand{\Oms}          {\ensuremath{\Omega^{\pm}}\xspace}
\newcommand{\degree}       {\ensuremath{^{\rm o}}\xspace}
\newcommand{\coaljet}      {\ensuremath{B_{\mathrm{2}}^{\mathrm{jet}}}\xspace}
\newcommand{\coalUE}       {\ensuremath{B_{\mathrm{2}}^{\mathrm{UE}}}\xspace}

%% file: fa_2026-01-22_Opt_C.tex

The ALICE Collaboration would like to thank all its engineers and technicians for their invaluable contributions to the construction of the experiment and the CERN accelerator teams for the outstanding performance of the LHC complex.
The ALICE Collaboration gratefully acknowledges the resources and support provided by all Grid centres and the Worldwide LHC Computing Grid (WLCG) collaboration.
The ALICE Collaboration acknowledges the following funding agencies for their support in building and running the ALICE detector:
A. I. Alikhanyan National Science Laboratory (Yerevan Physics Institute) Foundation (ANSL), State Committee of Science and World Federation of Scientists (WFS), Armenia;
Austrian Academy of Sciences, Austrian Science Fund (FWF): [M 2467-N36] and Nationalstiftung f\"{u}r Forschung, Technologie und Entwicklung, Austria;
Ministry of Communications and High Technologies, National Nuclear Research Center, Azerbaijan;
Rede Nacional de Física de Altas Energias (Renafae), Financiadora de Estudos e Projetos (Finep), Funda\c{c}\~{a}o de Amparo \`{a} Pesquisa do Estado de S\~{a}o Paulo (FAPESP) and The Sao Paulo Research Foundation  (FAPESP), Brazil;
Bulgarian Ministry of Education and Science, within the National Roadmap for Research Infrastructures 2020-2027 (object CERN), Bulgaria;
Ministry of Education of China (MOEC) , Ministry of Science \& Technology of China (MSTC) and National Natural Science Foundation of China (NSFC), China;
Ministry of Science and Education and Croatian Science Foundation, Croatia;
Centro de Aplicaciones Tecnol\'{o}gicas y Desarrollo Nuclear (CEADEN), Cubaenerg\'{\i}a, Cuba;
Ministry of Education, Youth and Sports of the Czech Republic, Czech Republic;
The Danish Council for Independent Research | Natural Sciences, the VILLUM FONDEN and Danish National Research Foundation (DNRF), Denmark;
Helsinki Institute of Physics (HIP), Finland;
Commissariat \`{a} l'Energie Atomique (CEA) and Institut National de Physique Nucl\'{e}aire et de Physique des Particules (IN2P3) and Centre National de la Recherche Scientifique (CNRS), France;
Bundesministerium f\"{u}r Forschung, Technologie und Raumfahrt (BMFTR) and GSI Helmholtzzentrum f\"{u}r Schwerionenforschung GmbH, Germany;
National Research, Development and Innovation Office, Hungary;
Department of Atomic Energy Government of India (DAE), Department of Science and Technology, Government of India (DST), University Grants Commission, Government of India (UGC) and Council of Scientific and Industrial Research (CSIR), India;
National Research and Innovation Agency - BRIN, Indonesia;
Istituto Nazionale di Fisica Nucleare (INFN), Italy;
Japanese Ministry of Education, Culture, Sports, Science and Technology (MEXT) and Japan Society for the Promotion of Science (JSPS) KAKENHI, Japan;
Consejo Nacional de Ciencia (CONACYT) y Tecnolog\'{i}a, through Fondo de Cooperaci\'{o}n Internacional en Ciencia y Tecnolog\'{i}a (FONCICYT) and Direcci\'{o}n General de Asuntos del Personal Academico (DGAPA), Mexico;
Nederlandse Organisatie voor Wetenschappelijk Onderzoek (NWO), Netherlands;
The Research Council of Norway, Norway;
Pontificia Universidad Cat\'{o}lica del Per\'{u}, Peru;
Ministry of Science and Higher Education, National Science Centre and WUT ID-UB, Poland;
Korea Institute of Science and Technology Information and National Research Foundation of Korea (NRF), Republic of Korea;
Ministry of Education and Scientific Research, Institute of Atomic Physics, Ministry of Research and Innovation and Institute of Atomic Physics and Universitatea Nationala de Stiinta si Tehnologie Politehnica Bucuresti, Romania;
Ministerstvo skolstva, vyskumu, vyvoja a mladeze SR, Slovakia;
National Research Foundation of South Africa, South Africa;
Swedish Research Council (VR) and Knut \& Alice Wallenberg Foundation (KAW), Sweden;
European Organization for Nuclear Research, Switzerland;
Suranaree University of Technology (SUT), National Science and Technology Development Agency (NSTDA) and National Science, Research and Innovation Fund (NSRF via PMU-B B05F650021), Thailand;
Turkish Energy, Nuclear and Mineral Research Agency (TENMAK), Turkey;
National Academy of  Sciences of Ukraine, Ukraine;
Science and Technology Facilities Council (STFC), United Kingdom;
National Science Foundation of the United States of America (NSF) and United States Department of Energy, Office of Nuclear Physics (DOE NP), United States of America.
In addition, individual groups or members have received support from:
FORTE project, reg.\ no.\ CZ.02.01.01/00/22\_008/0004632, Czech Republic, co-funded by the European Union, Czech Republic;
European Research Council (grant no. 950692), European Union;
Deutsche Forschungs Gemeinschaft (DFG, German Research Foundation) ``Neutrinos and Dark Matter in Astro- and Particle Physics'' (grant no. SFB 1258), Germany;
FAIR - Future Artificial Intelligence Research, funded by the NextGenerationEU program (Italy).

%% file: Alice_Authorlist_2026-01-22_Opt_C.tex
\begin{flushleft} 
\small

D.A.H.~Abdallah\,\orcidlink{0000-0003-4768-2718}\,$^{\rm 134}$, 
I.J.~Abualrob\,\orcidlink{0009-0005-3519-5631}\,$^{\rm 112}$, 
S.~Acharya\,\orcidlink{0000-0002-9213-5329}\,$^{\rm 49}$, 
G.~Aglieri Rinella\,\orcidlink{0000-0002-9611-3696}\,$^{\rm 32}$, 
L.~Aglietta\,\orcidlink{0009-0003-0763-6802}\,$^{\rm 24}$, 
N.~Agrawal\,\orcidlink{0000-0003-0348-9836}\,$^{\rm 25}$, 
Z.~Ahammed\,\orcidlink{0000-0001-5241-7412}\,$^{\rm 132}$, 
S.~Ahmad\,\orcidlink{0000-0003-0497-5705}\,$^{\rm 15}$, 
I.~Ahuja\,\orcidlink{0000-0002-4417-1392}\,$^{\rm 36}$, 
Z.~Akbar$^{\rm 79}$
% ...

\section*{Affiliation Notes}

$^{\rm I}$ Deceased\\
$^{\rm II}$ Also at: Max-Planck-Institut fur Physik, Munich, Germany\\
$^{\rm III}$ Also at: Czech Technical University in Prague (CZ)\\
$^{\rm IV}$ Also at: Instituto de Fisica da Universidade de Sao Paulo\\
$^{\rm V}$ Also at: Dipartimento DET del Politecnico di Torino, Turin, Italy\\
$^{\rm VI}$ Also at: Department of Applied Physics, Aligarh Muslim University, Aligarh, India\\
$^{\rm VII}$ Also at: Institute of Theoretical Physics, University of Wroclaw, Poland\\
$^{\rm VIII}$ Also at: Facultad de Ciencias, Universidad Nacional Aut\'{o}noma de M\'{e}xico, Mexico City, Mexico\\

\section*{Collaboration Institutes}

$^{1}$ A.I. Alikhanyan National Science Laboratory (Yerevan Physics Institute) Foundation, Yerevan, Armenia\\
$^{2}$ AGH University of Krakow, Cracow, Poland\\
$^{3}$ Bogolyubov Institute for Theoretical Physics, National Academy of Sciences of Ukraine, Kyiv, Ukraine\\
$^{4}$ Bose Institute, Department of Physics  and Centre for Astroparticle Physics and Space Science (CAPSS), Kolkata, India\\
$^{5}$ California Polytechnic State University, San Luis Obispo, California, United States\\
$^{6}$ Central China Normal University, Wuhan, China\\
$^{7}$ Centro de Aplicaciones Tecnol\'{o}gicas y Desarrollo Nuclear (CEADEN), Havana, Cuba\\
$^{8}$ Centro de Investigaci\'{o}n y de Estudios Avanzados (CINVESTAV), Mexico City and M\'{e}rida, Mexico\\
$^{9}$ Chicago State University, Chicago, Illinois, United States\\
$^{10}$ China Nuclear Data Center, China Institute of Atomic Energy, Beijing, China\\
$^{11}$ China University of Geosciences, Wuhan, China\\
$^{12}$ Chungbuk National University, Cheongju, Republic of Korea\\
$^{13}$ Comenius University Bratislava, Faculty of Mathematics, Physics and Informatics, Bratislava, Slovak Republic\\
$^{14}$ Creighton University, Omaha, Nebraska, United States\\
$^{15}$ Department of Physics, Aligarh Muslim University, Aligarh, India\\
$^{16}$ Department of Physics, Pusan National University, Pusan, Republic of Korea\\
$^{17}$ Department of Physics, Sejong University, Seoul, Republic of Korea\\
$^{18}$ Department of Physics, University of California, Berkeley, California, United States\\
$^{19}$ Department of Physics, University of Oslo, Oslo, Norway\\
$^{20}$ Department of Physics and Technology, University of Bergen, Bergen, Norway\\
$^{21}$ Dipartimento di Fisica, Universit\`{a} di Pavia, Pavia, Italy\\
$^{22}$ Dipartimento di Fisica dell'Universit\`{a} and Sezione INFN, Cagliari, Italy\\
$^{23}$ Dipartimento di Fisica dell'Universit\`{a} and Sezione INFN, Trieste, Italy\\
$^{24}$ Dipartimento di Fisica dell'Universit\`{a} and Sezione INFN, Turin, Italy\\
$^{25}$ Dipartimento di Fisica e Astronomia dell'Universit\`{a} and Sezione INFN, Bologna, Italy\\
$^{26}$ Dipartimento di Fisica e Astronomia dell'Universit\`{a} and Sezione INFN, Catania, Italy\\
$^{27}$ Dipartimento di Fisica e Astronomia dell'Universit\`{a} and Sezione INFN, Padova, Italy\\
$^{28}$ Dipartimento di Fisica `E.R.~Caianiello' dell'Universit\`{a} and Gruppo Collegato INFN, Salerno, Italy\\
$^{29}$ Dipartimento DISAT del Politecnico and Sezione INFN, Turin, Italy\\
$^{30}$ Dipartimento di Scienze MIFT, Universit\`{a} di Messina, Messina, Italy\\
$^{31}$ Dipartimento Interateneo di Fisica `M.~Merlin' and Sezione INFN, Bari, Italy\\
$^{32}$ European Organization for Nuclear Research (CERN), Geneva, Switzerland\\
$^{33}$ Faculty of Electrical Engineering, Mechanical Engineering and Naval Architecture, University of Split, Split, Croatia\\
$^{34}$ Faculty of Nuclear Sciences and Physical Engineering, Czech Technical University in Prague, Prague, Czech Republic\\
$^{35}$ Faculty of Physics, Sofia University, Sofia, Bulgaria\\
$^{36}$ Faculty of Science, P.J.~\v{S}af\'{a}rik University, Ko\v{s}ice, Slovak Republic\\
$^{37}$ Faculty of Technology, Environmental and Social Sciences, Bergen, Norway\\
$^{38}$ Frankfurt Institute for Advanced Studies, Johann Wolfgang Goethe-Universit\"{a}t Frankfurt, Frankfurt, Germany\\
$^{39}$ Fudan University, Shanghai, China\\
$^{40}$ Gauhati University, Department of Physics, Guwahati, India\\
$^{41}$ Helmholtz-Institut f\"{u}r Strahlen- und Kernphysik, Rheinische Friedrich-Wilhelms-Universit\"{a}t Bonn, Bonn, Germany\\
$^{42}$ Helsinki Institute of Physics (HIP), Helsinki, Finland\\
$^{43}$ High Energy Physics Group,  Universidad Aut\'{o}noma de Puebla, Puebla, Mexico\\
$^{44}$ Horia Hulubei National Institute of Physics and Nuclear Engineering, Bucharest, Romania\\
$^{45}$ HUN-REN Wigner Research Centre for Physics, Budapest, Hungary\\
$^{46}$ Indian Institute of Technology Bombay (IIT), Mumbai, India\\
$^{47}$ Indian Institute of Technology Indore, Indore, India\\
$^{48}$ INFN, Laboratori Nazionali di Frascati, Frascati, Italy\\
$^{49}$ INFN, Sezione di Bari, Bari, Italy\\
$^{50}$ INFN, Sezione di Bologna, Bologna, Italy\\
$^{51}$ INFN, Sezione di Cagliari, Cagliari, Italy\\
$^{52}$ INFN, Sezione di Catania, Catania, Italy\\
$^{53}$ INFN, Sezione di Padova, Padova, Italy\\
$^{54}$ INFN, Sezione di Pavia, Pavia, Italy\\
$^{55}$ INFN, Sezione di Torino, Turin, Italy\\
$^{56}$ INFN, Sezione di Trieste, Trieste, Italy\\
$^{57}$ Inha University, Incheon, Republic of Korea\\
$^{58}$ Institute for Gravitational and Subatomic Physics (GRASP), Utrecht University/Nikhef, Utrecht, Netherlands\\
$^{59}$ Institute of Experimental Physics, Slovak Academy of Sciences, Ko\v{s}ice, Slovak Republic\\
$^{60}$ Institute of Physics, Homi Bhabha National Institute, Bhubaneswar, India\\
$^{61}$ Institute of Physics of the Czech Academy of Sciences, Prague, Czech Republic\\
$^{62}$ Institute of Space Science (ISS), Bucharest, Romania\\
$^{63}$ Institut f\"{u}r Kernphysik, Johann Wolfgang Goethe-Universit\"{a}t Frankfurt, Frankfurt, Germany\\
$^{64}$ Instituto de Ciencias Nucleares, Universidad Nacional Aut\'{o}noma de M\'{e}xico, Mexico City, Mexico\\
$^{65}$ Instituto de F\'{i}sica, Universidade Federal do Rio Grande do Sul (UFRGS), Porto Alegre, Brazil\\
$^{66}$ Instituto de F\'{\i}sica, Universidad Nacional Aut\'{o}noma de M\'{e}xico, Mexico City, Mexico\\
$^{67}$ iThemba LABS, National Research Foundation, Somerset West, South Africa\\
$^{68}$ Jeonbuk National University, Jeonju, Republic of Korea\\
$^{69}$ Korea Institute of Science and Technology Information, Daejeon, Republic of Korea\\
$^{70}$ Laboratoire de Physique Subatomique et de Cosmologie, Universit\'{e} Grenoble-Alpes, CNRS-IN2P3, Grenoble, France\\
$^{71}$ Lawrence Berkeley National Laboratory, Berkeley, California, United States\\
$^{72}$ Lund University Department of Physics, Division of Particle Physics, Lund, Sweden\\
$^{73}$ Marietta Blau Institute, Vienna, Austria\\
$^{74}$ Nagasaki Institute of Applied Science, Nagasaki, Japan\\
$^{75}$ Nara Women{'}s University (NWU), Nara, Japan\\
$^{76}$ National Centre for Nuclear Research, Warsaw, Poland\\
$^{77}$ National Institute of Science Education and Research, Homi Bhabha National Institute, Jatni, India\\
$^{78}$ National Nuclear Research Center, Baku, Azerbaijan\\
$^{79}$ National Research and Innovation Agency - BRIN, Jakarta, Indonesia\\
$^{80}$ Niels Bohr Institute, University of Copenhagen, Copenhagen, Denmark\\
$^{81}$ Nikhef, National institute for subatomic physics, Amsterdam, Netherlands\\
$^{82}$ Nuclear Physics Group, STFC Daresbury Laboratory, Daresbury, United Kingdom\\
$^{83}$ Nuclear Physics Institute of the Czech Academy of Sciences, Husinec-\v{R}e\v{z}, Czech Republic\\
$^{84}$ Oak Ridge National Laboratory, Oak Ridge, Tennessee, United States\\
$^{85}$ Ohio State University, Columbus, Ohio, United States\\
$^{86}$ Physics department, Faculty of science, University of Zagreb, Zagreb, Croatia\\
$^{87}$ Physics Department, Panjab University, Chandigarh, India\\
$^{88}$ Physics Department, University of Jammu, Jammu, India\\
$^{89}$ Physics Program and International Institute for Sustainability with Knotted Chiral Meta Matter (WPI-SKCM$^{2}$), Hiroshima University, Hiroshima, Japan\\
$^{90}$ Physikalisches Institut, Eberhard-Karls-Universit\"{a}t T\"{u}bingen, T\"{u}bingen, Germany\\
$^{91}$ Physikalisches Institut, Ruprecht-Karls-Universit\"{a}t Heidelberg, Heidelberg, Germany\\
$^{92}$ Physik Department, Technische Universit\"{a}t M\"{u}nchen, Munich, Germany\\
$^{93}$ Politecnico di Bari and Sezione INFN, Bari, Italy\\
$^{94}$ Research Division and ExtreMe Matter Institute EMMI, GSI Helmholtzzentrum f\"ur Schwerionenforschung GmbH, Darmstadt, Germany\\
$^{95}$ Saga University, Saga, Japan\\
$^{96}$ Saha Institute of Nuclear Physics, Homi Bhabha National Institute, Kolkata, India\\
$^{97}$ School of Physics and Astronomy, University of Birmingham, Birmingham, United Kingdom\\
$^{98}$ Secci\'{o}n F\'{\i}sica, Departamento de Ciencias, Pontificia Universidad Cat\'{o}lica del Per\'{u}, Lima, Peru\\
$^{99}$ SUBATECH, IMT Atlantique, Nantes Universit\'{e}, CNRS-IN2P3, Nantes, France\\
$^{100}$ Sungkyunkwan University, Suwon City, Republic of Korea\\
$^{101}$ Suranaree University of Technology, Nakhon Ratchasima, Thailand\\
$^{102}$ Technical University of Ko\v{s}ice, Ko\v{s}ice, Slovak Republic\\
$^{103}$ The Henryk Niewodniczanski Institute of Nuclear Physics, Polish Academy of Sciences, Cracow, Poland\\
$^{104}$ The University of Texas at Austin, Austin, Texas, United States\\
$^{105}$ Universidad Aut\'{o}noma de Sinaloa, Culiac\'{a}n, Mexico\\
$^{106}$ Universidade de S\~{a}o Paulo (USP), S\~{a}o Paulo, Brazil\\
$^{107}$ Universidade Estadual de Campinas (UNICAMP), Campinas, Brazil\\
$^{108}$ Universidade Federal do ABC, Santo Andre, Brazil\\
$^{109}$ Universitatea Nationala de Stiinta si Tehnologie Politehnica Bucuresti, Bucharest, Romania\\
$^{110}$ University of Cape Town, Cape Town, South Africa\\
$^{111}$ University of Derby, Derby, United Kingdom\\
$^{112}$ University of Houston, Houston, Texas, United States\\
$^{113}$ University of Jyv\"{a}skyl\"{a}, Jyv\"{a}skyl\"{a}, Finland\\
$^{114}$ University of Kansas, Lawrence, Kansas, United States\\
$^{115}$ University of Liverpool, Liverpool, United Kingdom\\
$^{116}$ University of Science and Technology of China, Hefei, China\\
$^{117}$ University of Silesia in Katowice, Katowice, Poland\\
$^{118}$ University of South-Eastern Norway, Kongsberg, Norway\\
$^{119}$ University of Tennessee, Knoxville, Tennessee, United States\\
$^{120}$ University of the Witwatersrand, Johannesburg, South Africa\\
$^{121}$ University of Tokyo, Tokyo, Japan\\
$^{122}$ University of Tsukuba, Tsukuba, Japan\\
$^{123}$ Universit\"{a}t M\"{u}nster, Institut f\"{u}r Kernphysik, M\"{u}nster, Germany\\
$^{124}$ Universit\'{e} Clermont Auvergne, CNRS/IN2P3, LPC, Clermont-Ferrand, France\\
$^{125}$ Universit\'{e} de Lyon, CNRS/IN2P3, Institut de Physique des 2 Infinis de Lyon, Lyon, France\\
$^{126}$ Universit\'{e} de Strasbourg, CNRS, IPHC UMR 7178, F-67000 Strasbourg, France, Strasbourg, France\\
$^{127}$ Universit\'{e} Paris-Saclay, Centre d'Etudes de Saclay (CEA), IRFU, D\'{e}partment de Physique Nucl\'{e}aire (DPhN), Saclay, France\\
$^{128}$ Universit\'{e}  Paris-Saclay, CNRS/IN2P3, IJCLab, Orsay, France\\
$^{129}$ Universit\`{a} degli Studi di Foggia, Foggia, Italy\\
$^{130}$ Universit\`{a} del Piemonte Orientale, Vercelli, Italy\\
$^{131}$ Universit\`{a} di Brescia, Brescia, Italy\\
$^{132}$ Variable Energy Cyclotron Centre, Homi Bhabha National Institute, Kolkata, India\\
$^{133}$ Warsaw University of Technology, Warsaw, Poland\\
$^{134}$ Wayne State University, Detroit, Michigan, United States\\
$^{135}$ Yale University, New Haven, Connecticut, United States\\
$^{136}$ Yildiz Technical University, Istanbul, Turkey\\
$^{137}$ Yonsei University, Seoul, Republic of Korea\\
$^{138}$ Affiliated with an institute formerly covered by a cooperation agreement with CERN\\
$^{139}$ Affiliated with an international laboratory covered by a cooperation agreement with CERN.\\

\end{flushleft} 